# End-to-End Architecture Modularisation and Slicing for Next Generation Networks

Xueli An, Riccardo Trivisonno, Huawei Technologies

Hans Einsiedler, Dirk von Hugo, Kay Haensge, Deutsche Telecom

Xiaofeng Huang, Qing Shen, Orange

Daniel Corujo, IT Aveiro

Kashif Mahmood, Telenor

Dirk Trossen, Interdigital

Marco Liebsch, Filipe Leitao, NEC

Cao-Thanh Phan, Frédéric Klamm, BCOM

### Abstract

The journey towards the deployment of next generation networks has recently accelerated, driven by the joint effort of research and standards organisations. Despite this fact, the overall picture is still unclear as prioritization and understanding on several key concepts are not yet agreed by major vendors and network providers. Network Slicing is one of the central topics of the debate, and it is expected to become the key feature of next generation networks, providing the flexibility required to support the variety of 5G use cases and business. Network slices are seen as network operator business, offering the possibility to provide flexible services and even infrastructures to vertical industries and classical Telco customers alike. Another key ingredient is the Architecture Modularisation concept, discussed in this paper and regarded by the authors as the essential design principle to build a flexible network architecture natively supporting Network Slicing. According to this concept, conventional monolithic network functions, often corresponding to physical network elements in the existing systems, are to split into basic building blocks defined with the proper granularity, allowing the definition of different logical architectures (i.e. different Network Slices). In this paper, we further discuss a modularisation methodology as a criteria to define the 'right' set of basic building blocks. Defined through this proposed methodology, the set of basic building blocks and the relating interfacing model are discussed. The paper concludes by proposing a modular 5G network architecture as candidate for next generation network standards.

## 1 INTRODUCTION

Triggered and carried out by several (world-wide) ongoing 5G research projects (e.g. [1], [2], and [3]), and already formalised by standardisation organisations (e.g. [4]), the analysis of 5G use cases becomes mature, and they are grouped into three main categories: enhanced Mobile Broadband (eMBB), Massive Internet of Things (mIoT), and Critical Communications. Despite the time and effort dedicated to this analysis, an objective technical justification for the need of a new generation mobile network fails to emerge, as several use cases are seemingly supported by a smart re-engineering of 4G systems (e.g. [6], [7]). However, the plethora of different, partly incompatible options specified today within the 4G architecture is increasingly considered as a burden: instead of providing the flexibility, it forces operators to make either-or decisions, while the standardization bodies spend tremendous efforts to maintain the overall patchwork that 4G has evolved into over years. A cleaner, simpler approach is therefore sought for, and the only driver all actors seem to agree on is the need for architecture flexibility, allowing the integration of upcoming services and new business actors of the verticals and customer segment within the Telco domain, although no concrete technical solutions satisfy network operator expectations so far. Still defined at a quite high level, the concept of Network Slicing may be the answer for satisfying the need of a plurality of coexisting network architectures, each slice tailored to different functional and performance requirements, potentially operating isolated from each another, and instantiated over a common virtual or physical infrastructure. In this landscape, this paper aims at taking a significant and concrete step forward, proposing a methodology for architecture modularisation to be able to implement such slicing. The formalised methodology aims at defining the basic set of network Building Blocks (BBs), upon which access agnostic End to End (E2E) architectures can be defined. This methodology focuses on Core Network (CN) functional blocks, and defines Access Network (AN) abstractions to be integrated in the E2E architecture. Moreover, the defined set of BBs needs to allow tailored network architectures to be instantiated as independent Network Slices, yet enabling 5G devices to discover and connect to multiple network slices, and allowing inter-slice operations and optimisation for use cases requiring it.

The paper is organised as follows. Consolidating the results of the multilateral and industry-driven CONFIG project [3], section 2 summarises the 5G architecture cornerstones and the reference model, which are assumed as starting point for this work. Section 3 proposes the modularisation methodology, aimed at designing the key building blocks and the interfaces defined by the reference model. Section 4 presents the high level design of building blocks as well as the resulting overall 5G network architecture. Finally, Section 5 and 6 highlight the open issues requiring further investigation and summarise the key conclusions of the paper.

## 2  5G Architecture Cornerstones

The attempt to define novel architectures for 5G systems has been undertaken by several joint research initiatives, sometimes with overlapping scope, coherent goals and similar achievements. A relevant example is 5G Norma [2], which aims at developing a software-controlled E2E 5G mobile network architecture enabling multi-service, context-aware adaptation of network functions, and supporting usage of multiple infrastructures (multi-tenancy), while focusing on cellular radio systems. Similarly, Metis II [1] aims at an overall 5G radio AN design and at providing technical enablers to integrate 5G technologies and components currently being developed. Beyond the scope of cellular radio systems, the CONFIG project [3] is dedicated to the design of a 5G convergent CN architecture with the goal to achieve architecture flexibility, support heterogeneous access, and allow for vertical business integration, leveraging on recent advances on Network Function Virtualization (NFV), Software Defined Networking (SDN). Focusing on the CN, CONFIG laid the foundations of an E2E 5G system which embeds the abstraction for AN.

The initial results, delivered in [5], are briefly summarised in the following. To enable the design of logical architectures tailored to performance and functional requirements of different use cases, the principle of architecture modularisation via network function decomposition was adopted: the conventional monolithic network functions (often corresponding to physical network elements in 4G), both for the

Control-plane (C-plane) and Data-plane (D-plane), are to be split into *Building Blocks* (BBs) defined with the proper granularity, thus allowing for defining different logical architectures via the interconnection of different subsets of C-plane and D-plane BBs.

In the process of decomposing the monolithic network functions into BBs, the distinction between BBs relating to the *Access Network* (AN) and the *Core Network* (CN) emerged. To minimise their dependency and achieve the definition of a Convergent network, providing connectivity via a multitude of accesses (not only including cellular radio) via a common CN, a different AN-CN functional split and interface model are necessary.

Besides flexibility, Architecture Modularisation provides the essentials to enable network slicing, as a network slice is defined as an independent logical network, made by the interconnection of a subset of required BBs, and which can be independently instantiated and operated over physical or virtual infrastructure. The network slice is defined across the whole communication system, hence including both AN and CN functions and the respective nodes including the end-systems (e.g. mobile terminals, customer premises equipment), and may involve last hop connection resources, including spectrum, radio resources, processing, storage, etc. The implementation of this concept requires tackling three key issues: design, instantiation, and operation. Designing a slice comprises the definition of required network functions associated with the C-/D-plane architecture, procedures, and protocols upon the basic set of BBs. Lifecycle management comprises the instantiation of the slice's BBs in the physical/virtual infrastructure, as well as the operation of the slice in terms of configuration, management, and monitoring mechanisms.

A support of network slicing would be achieved by creating a toolbox of BBs and relating interfacing protocols allowing the implementation (i.e. design and instantiation) of specific network slices with the flexibility of SDN and NFV technologies. From the point of view of an operator and per-slice, two kinds of toolboxes will exist. The operator internal toolbox – operator repository –, which contains its standardised network functions and protocols. Second, an external toolbox, containing use case specific network functions with standardised interfaces, allowing e.g. the integration of verticals which might be providing customised and proprietary BBs.

The CONFIG project's initial phase concluded with the formalisation of a reference model for Architecture Modularisation and network slicing illustrated in Figure 1, which has been discussed and approved in 3GPP SA2 Study Item [5]. The reference model formalises the BBs as founding logical elements for different architectures and network slices, the separation between C- and D-planes and between the AN and CN. Also, the model defines the reference points, including interfaces in between BBs, between C- and D-plane, as well as a West Bound Interface (WBI), which allows C-plane functions in the CN and the AN, up to the UE, to exchange information. The model also highlights each BB as composed by a set of Sub-functions (SFs) by which BBs customisation can be achieved.

The next step requires the definition and the design of the set of BBs and related interfaces, which is a highly complex engineering task, essentially for two reasons. First, the problem has many variables and few constraints and, hence, a wide solution space. Second, there are no clear and objective metrics to evaluate and compare alternative architectures. Often, novel network architectures are designed as incremental evolution of existing ones, in the attempt of ensuring systems backward compatibility and to fix specific shortcomings, but this approach does not fit the need of defining a clean-slate architecture. The following section elaborates a modularisation methodology for the definition of BBs and related interfaces.

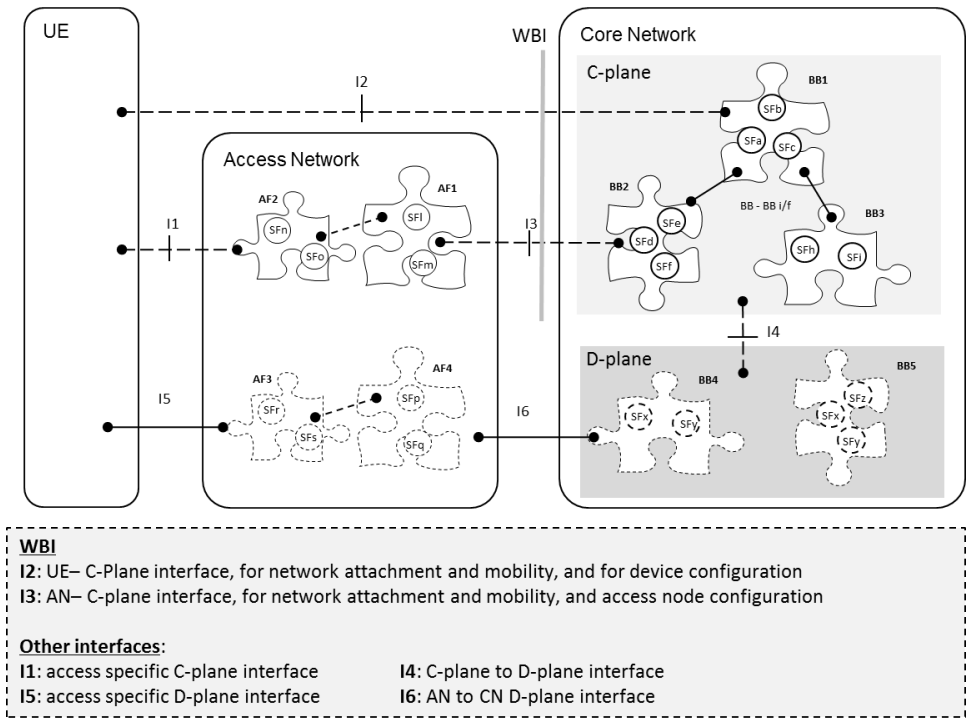

*Figure 1: Architecture Modularization Reference Model*

# 3 MODULARIZATION METHODOLOGY

A **BB** is a well-scoped network function, dedicated to perform specific C-/D-plane tasks and accessible via interfaces. Any modularization method to define BBs needs to consider the trade-off between *flexibility* and *complexity*, because larger number of elementary modules would provide higher flexibility, while the additional interface/interconnection complexity would pose implementation challenges.

A balance between these conflicting requirements can be achieved by grouping homogeneous network functional areas in wider scope BBs (to avoid a too fine granularity) and to further design each BB as the composition of elementary *Sub-Functions* (SF).

Different customised versions of BBs can be defined via a proper composition of different elementary SFs. Different interconnections of different sets of BBs define different Logical Architectures fulfilling requirements of different use cases. The methodology, seeking the proper SFs grouping into BBs, is made by four steps, illustrated in Figure 2.

The first step identifies the necessary SFs based on the fundamental services provided by the CN such as session management, mobility management, etc. Moreover, the required SFs can also be identified based on 5G specific features, e.g. support network slicing, as well as specific requirements from use cases, e.g. seamless mobility support for vehicle to infrastructure (V2I) communication, etc.

The second step identifies the SFs which need to be kept separated from each other, and hence assimilated to different BBs. The criteria for SFs separation are:

1. **SF placement**: some SFs need to be instantiated in proximity to ANs (edge SFs), e.g. to reduce the C-plane and D-plane latency; in contrast, other SFs will need to be centrally instantiated (core SFs), e.g. to obtain a system-level view to execute certain operations.
2. **SF re-usability**: SFs that can be potentially used by multiple service layers or have different internal variants shall be separated from SFs that are specific for certain services.
3. **SF optionality**: a use case specific SF should be kept separated from those that will be required by all or multiple use cases. For instance, the policy control SF may or may not be necessary for devices connecting through a fixed access, while handover management SF will not be used for fixed network usages.
4. **SF evolution cycle**: a rapidly evolving SF shall be kept separated from those that have a slower evolution cycle. As an example, the authentication function may evolve rapidly and independently from the others because of the introduction of new authentication methods.

The third step deals SFs not having separation constraints and determines, how they can be combined into BBs. As a fundamental rule, BBs shall combine SFs from the same functional domain, as defined by legacy network (e.g. mobility, security, charging, etc.). Further combination guideline is the reduction of inter-BB interfacing complexity. When SFs are combined, implementation shall grant to each of them an independent scalability within the BB.

The fourth step consists of refining the BBs and SFs. BB refinement and SF redefinition may occur, for instance, if the defined BBs result in an inefficient information exchange during pre-defined procedures. In such a case, it will move back to step 1 or step 3 according to the results from step 4. Otherwise, the definition of BB is complete.

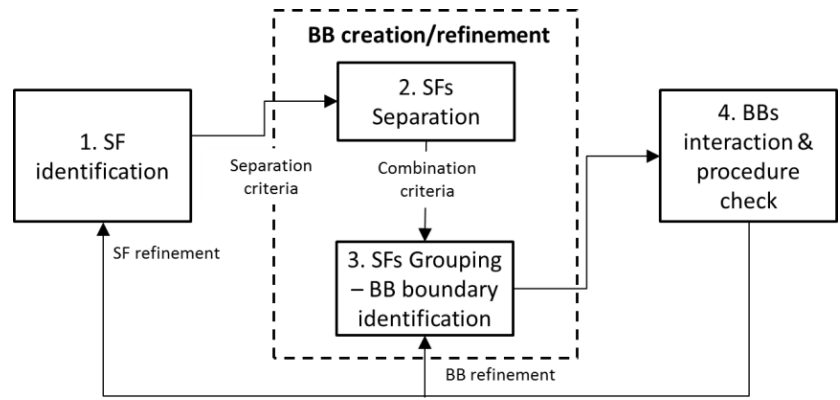

*Figure 2: Modularization Methodology Flowchart*

## 4 BUILDING BLOCKS AND INTERFACE MODEL

The definition of BBs follows the modularisation methodology per Section 3 and considers the full list of elementary network functions of 4G networks, plus additional ones which are proposed as 5G differentiators. Table 1 summarises the 4G network function analysis and the grouping to define BBs. The table is derived from an in-depth analysis of 3GPP specifications [8]. The 1st and 2nd columns summarise the proposed BBs and their corresponding SFs. The 3rd column describes the role of an SF, and the 4th indicates the originator of the SF, which may come from 3GPP or from 5G specific features. The key BBs are further explained in the following sub-sections.

*Table 1: BB and Sub-Function Definition*

| BB | Sub-function (SF) | SF Identification | |
|---|---|---|---|
| | | SF Description | Originator |
| Access Function (AF) | D-Plane Control | Control of AN D-plane | - |
| | AN management | Radio connection related management | 3GPP |
| | CN related access control | Identifying permissions for CP entity access/modification rights for AN configuration | - |
| | Path record | Forwarding path management at AN | 3GPP |
| Connectivity Management (CM) | Network access control and configuration | Admission and access control | 3GPP |
| | Access Functions Control | Handling the integration of alternative AN terminating AF specific interfaces (Convergence) | 5G specific |
| | Session management | Handling Session Establishment/Continuity/Release and well as Data Plane function selection. | 3GPP |
| | Slice management | Handling slice selection and attachment, inter-slice mobility and optimisation | 5G specific |
| | Roaming management | Handling Roaming and inter slice roaming | 3GPP |
| Mobility Management (MM) | Mobility policy enforcement | Handling different mobility policy for different use cases | 5G specific |
| | Device location tracking | Device location (e.g. tracking area) management | 3GPP |
| | Device paging | User reachability provisioning | 3GPP |
| | Mobility assistance | Supporting device-centric mobility for convergence purpose | 5G specific |
| Security and AAA Management (SAM) | Unified database for identity management | ID management for convergence purpose | 5G specific |
| | Authentication and authorization | Handling device/ UE Authentication and Authorization | 3GPP |
| | Single sign-on | Authentication and Authorization | 5G specific |
| | Security monitoring | Monitoring security related aspects of other C-Plane BBs as well as checking the integrity between the network entities. | 5G specific |
| Flow Management (FM) | Forwarding plane monitoring | Monitoring related to the data plane. | 3GPP |
| | Forwarding path definition | Defining the forwarding paths based on relevant metrics. | 5G specific |
| | Flow management decision | Handling of D-plane functions selection/reselection for flow control and management. | 3GPP |
| Context Generation and Handling Function (CGHF) | Pub Sub Framework management | Handling Publication and Subscription of/to information for Context Generation by BBs | 5G specific |
| | Context generation | Processing information collected from BBs and converting into Context Information | 5G specific |
| | Context Management | Describing the model to generate Context Information. | 5G specific |

### 4.1.1 Access Function (AF)

The AF integrates different SFs that are used to abstract the last-hop connectivity by supporting the WBI with abstractions for radio connection management (attachment, mobility and node configuration), independently of the access technology. Towards the AN, access-specific protocols are used, towards the

CN, an interface which is common to all ANs is established. The AF should also manifest itself as the converging point for CN-related access control management procedures, such as initial access, transition, or re-entry of end devices into the AN, towards the control plane Complementing these mechanisms, the AF also handles path record aspects for mobility- and nomadic-related scenarios, allowing controlling entities to be aware of the latest network end-point to which devices were connected.

### 4.1.2 Connectivity Management (CM)

The CM provides the interfaces towards the AN and 5G devices for Network Access Stratum (NAS) signalling. After a device attaches to the 5G network, CM selects D-plane functions (e.g. mobility anchor, PDN anchor) and manages session for this device. A device may maintain single or multiple-connectivity within one or more ANs and via one or multiple flows within the core simultaneously, hence CM maintains a device's convergent state machine and the corresponding context. It is responsible for dynamically allocating an identifier (e.g. IP address or other address schemes) for a device to access the network and use services. A device may require more than one such ID for different ANs. Moreover, CM can discover a device's capability and connectivity relevant parameters, and CM can suggest certain configurations for devices (e.g. using Wi-Fi other than LTE for offloading purposes). CM handles the connectivity-related procedures by virtue of a global view of a device and AN resource related usage information. Hence, it can also provide AF-related configuration parameters.

The CM is the key BB allowing per-slice customization of the C-plane. Figure 3 depicts two methods to enable slice selection at C-plane. For the first method, CM contains two parts: global part and local part. The global part performs first authentication of the attached device and selects an appropriated slice, e.g. to its subscription information, hence it is the common part among different slices. Afterwards, it attaches to the selected slice and uses a local CM, which is slice-specific. For the second method, CM is slice specific, a device may attach to a default slice, if this is not the appropriated slice for the device (e.g. not sufficient subscription information), the local CM may suggest another slice to attach for the device.

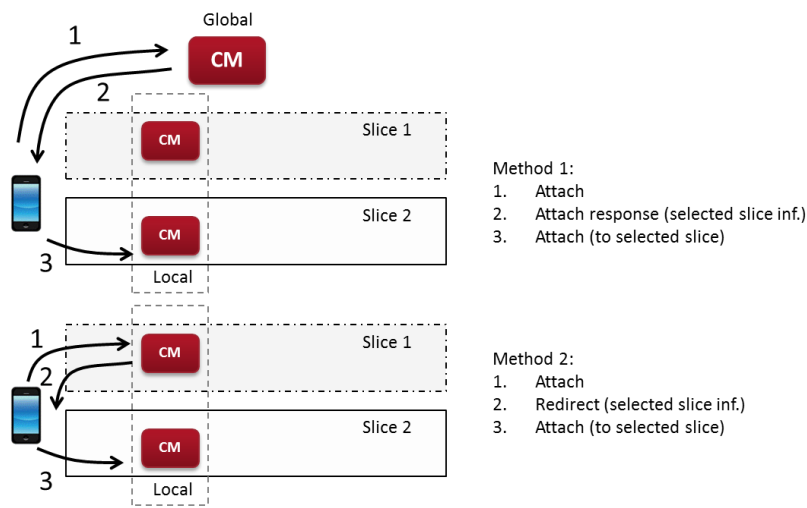

*Figure 3: CM supported slice selection methods*

### 4.1.3 Flow Management (FM)

The FM controls how D-plane flows are traversing the core between D-plane functions towards access domains and towards other domains such as servers hosting application. Moreover, the FM manages the related procedures which involve establishing and reconfiguring D-plane paths. FM also enforces QoS related policies, which includes resource reservation for critical flows and path management based on QoS profiles.

To remain independent of the particular D-plane technology, the FM utilises common semantics on its interface towards the transport network, which is also denoted as southbound interface (SBI), i.e. I4 in Figure 1, to have the full control to configure relevant rules on the D-Plane and to receive notifications from the D-Plane, e.g. this can be accomplished by an adaptor function to terminate the FM's SBI. This adaptor in turn may utilise a variety of protocols/interfaces to collaborate with the controller of the transport (e.g. SDN controller) and the associated applications in the underlying D-plane. Therefore, the FM can also enable per slice customised D-plane model and procedure. For instance, the FM can enforce shortest path route on slice 1, and optimised load distribution path on slice 2.

The FM has peering with a number of other BBs like CM, SAM, and AF, e.g. the device state information from the CM can be used by FM for flow optimization. Likewise, the SAM receives policies managed by the operator to deploy dynamically and pervasively security services in the FM. These services aim to enforce the protection of data flows as ciphering and integrity processing or of the network as flow packet inspection or firewalling. From an implementation perspective, the FM can have a distributed implementation which means it can be instantiated in multiple locations, while, from a business perspective FM offers the potential of being implemented in different ways with diverse features.

### 4.1.4 Mobility Management (MM)

The MM is the BB for deciding, initiating, and maintaining all major connection issues and the related active sessions which are impacted by a change in the connection endpoints. Mobility events causing such changes may origin from actual movement of the device, but also changes in propagation conditions or local AN load. Feedback on relevant changes in connection parameters can also be forwarded to the CN C-plane, and a decision including different access domains as well as a change in responsible slices can generate an action in the CN C-plane. The MM is in charge of tasks which will enable 5G system to extent currently-restricted mobility capabilities in 4G-EPC. Thus, in addition to control handover execution (across multiple access technologies or access domains) as well as to support device reachability related procedures, it will allow 5G system to feature: First, access agnostic mobility support, i.e. MM procedures shall be triggered towards/from AFs (e.g. via remote Mobility Assistant SF) for any access technology supported by 5G; second, seamless inter-AN mobility with session continuity (including session endpoints connected to multiple access nodes of same/different access technologies under same/different operators administrative domain); third, slice-specific mobility schemes (e.g. for unicast, multicast, or broadcast communications); fourth, mobility pattern-based schemes (e.g. state change for plurality of devices such as in public or individual transport/traffic scenario); fifth, control and monitoring of location and accessibility also for devices in idle state per Location Tracking/Paging SFs; sixth, host-based mobility across different domains enabled via Mobility Assistant SF placed at the UE.

The MM can be optional to enrich a slice with mobility support. Within slices not supporting mobility (e.g. the residential/fixed broadband access use case), the CM shall directly interact with AF without involving MM. Different mobility support policies shall apply to different slices, e.g. mobility support for V2I may be different from mobility support for eMBB. This allows for different types of (granular) MM to be developed and applied related to protocols of different complexity. The actual instantiation of MM can be mobility scenario dependent (e.g., depending on UE speed and environment activation of either multiple distributed or few centralised anchor nodes will provide best performance). Those decisions are taken by the Mobility Policy Enforcement SF.

### 4.1.5 Security and AAA Management (SAM)

The SAM is responsible to perform authentication and security related functions. It may evolve rapidly and independently because of the introduction of new authentication and security methods which is quite different from other BBs. It is the integral component for the whole authentication and authorisation procedures even when a low-secure authentication scheme is used.

The SAM provides the mutual authentication and key agreement of the device and the network. It generates a security context and keys. The latter are used in order to protect the signalling and user data flows integrity and confidentiality. This feature prevents the impersonation of the originator, changes to the content and unauthorised wire-tapping. In addition, the SAM protects the permanent subscriber identity from disclosure by means of randomised pseudonyms or other, adequate mechanisms. The AAA functionality shall be access-agnostic in a 5G control plane. Hence, the SAM should support generalised Authentication and Authorization protocols (in terms of Authenticator and Single Sign-On SFs) for the connecting device as well as providing an interface for other services to enable a trustful E2E relationship between the device, the network and the accessed service.

The wide range of device features influence security. Their types (e.g. smart phone, low cost sensor) imply different capabilities. Their origins (e.g. personal, company issued) challenge trust models. Their business verticals (e.g. industry, health, vehicle …) impose proprietary standards. This diversity of security functions requirements is hard and complex to satisfy. Consequently, a data flows requiring the same capabilities and/or security level, should be grouped and confined, e.g. using slicing. By means of offering tool sets and protocols allowing for granular AAA and security levels different policies can be applied to different use cases and slices, e.g. in mIoT slice for sensors only a low-secure authentication and authorization scenario may be required.

The Monitoring SF in addition logs critical events and transactions for incidents investigation and auditing purpose; therefore it provides the traceability and the accountability.

### 4.1.6 Context Generation and Handling Function (CGHF)

CGHF is responsible for generating, storing and distributing rich context which can be utilised by both D-plane and C-plane BBs which subscribe to it. The context refers to the information that describes the situation of an entity which can be a device, user, network, radio environment or an external application. The context which we refer to here is richer than what is available in today's network because it is composed of intelligent merge of input from various sources taking into account also the historical data. An example of a rich context that describes the situation of application is that "an application X is about to suffer more than normal latency". A possible action, that a BB (e.g. CM) can take if subscribed to this context, can be D-plane function reselection.

The CGHF provides this rich context in such a way that it adheres to the principle of modularization that would enable a cross-NF management of context information and would provide the desired reusability of context information. This would also provide an interface for exchange of contexts to 3$^{rd}$ parties. It needs to be highlighted that the context consumption and the triggering of the relevant actions happens inside the individual BBs and not in CGHF as the mission specific BBs are in the best position to take the corresponding actions.

The CGHF collects inputs from different sources such as UE, AN, CN and even 3rd party applications. One or multiple inputs are then converted into a context by applying the appropriate analysis model. Finally the contexts are fed back to the BBs or 3rd party applications.

The CGHF determines which and how information from the individual input is to be combined, which kind of analysis needs to be applied, which interfaces should be used for information exchange between other BBs, how to prevent unwanted share of information and how to deal with any conflict resolution among BBs using context.

## 4.2 INTERFACE MODEL

Two types of interfaces are considered in the architecture design. The first is the inter-BB interface, which allows interconnection and information exchange among BBs within CN, and the second is the WBI. C-plane signalling for network attachment and mobility may or may not involve the mediation of the AF. The UE may perform direct signalling with the CN C-Plane on interface I2. If AF mediation is needed, the UE performs access specific signalling on interface I1 and the AF handles C-Plane signalling on interface I3. The signalling may involve CN BBs and/or other AFs. Interface I3 is also meant for location tracking and UE reachability not establishing direct C-Plane signalling for those purposes, as well as for AF configuration.

The inter-BB interfaces within CN C-plane shall be specified or even standardised, to enable an efficient exchange of information and control messages between BBs developed and deployed by potentially different players. At the same time, a high flexibility shall govern the interface design to allow for further extensions of the above described BBs, as well as addition of potential new BBs, which may be required to serve the needs of upcoming, yet unknown, services and applications.

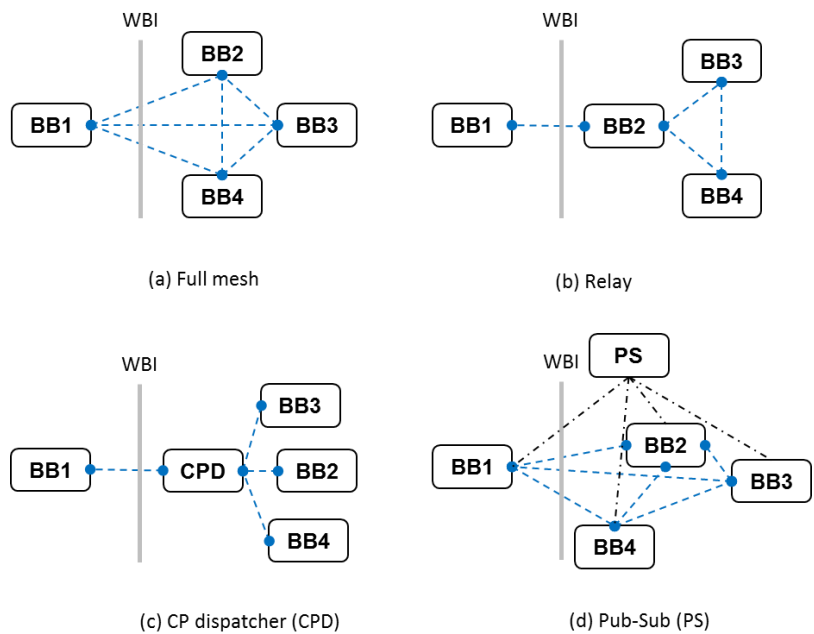

Figure 4: Interface models

Figure 4 illustrates interface models that can be applied:

a) Full mesh model: Each BB has interfaces towards all the relevant BBs.

b) Relay model: a single BB as termination of the WBI, with the additional task of relaying C-plane signalling to the concerned BBs in CN.
c) C-plane dispatcher (CPD) model: the CN C-plane signalling entry point dispatches the C-plane traffic to relevant BBs, the inter CN BB communication also relies on CPD. For this model, CPD acts as a proxy, and for each BB a set of proxy interfaces can be defined, which only provides the relevant information to the corresponding BB and its proxy interface. Higher flexibility towards inclusion of new BBs is achieved by using a common interconnection function, which all BBs have access to via a BB-specific interface.
d) Publish-subscribe (PS) model: this is the most dynamic model, which provides asynchronous inter-component communication [10] by BBs subscribing to messages submitted to a common network function (e.g. PS as shown in subplot d) in order to receive relevant information from those BBs providing it.

## 4.3 OVERALL 5G ARCHITECTURE

The BBs and interfaces design illustrated above allows evolving from the reference model depicted in Figure 1 to a proposal for a more complete E2E network architecture, shown in Figure 5. The architecture highlights BBs, inter BBs interfaces, and interfaces between BBs with the AN and the UE. The picture hints to an analogous modularisation approach applied to the D-Plane and shows C-plane and D-plane interfaces. It should be noted that the picture refers to the most complete case of a slice including all BBs defined in section 4.1 and to interface option (a) defined in section 4.2. As some BBs have been defined as "optional" building blocks (i.e. MM, CGHF), simplified architectures can be derived removing BBs when not required by use cases the network slice is designed for. Additionally, tailored architecture can be defined by further customisation of the BBs, which may include different sets of SFs.

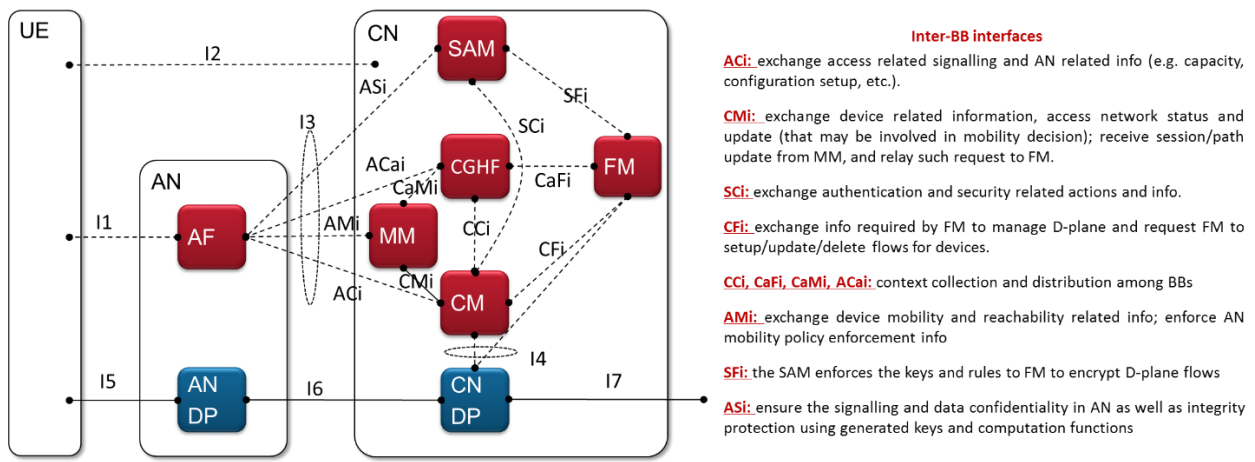

*Figure 5: Overall 5G End to End Architecture*

## 5 NEXT STEPS

The architecture depicted in Figure 5 is a valid candidate for adoption in next generation network standards. However, careful reviewers will not miss to identify several critical open issues to address in order to finalise the design.

First, the best design choice for WBI needs to be evaluated. As discussed in section 4, CM, MM, SAM and CGHF require information exchange with the AF. The choice illustrated in Figure 5 assumes that the AF is directly connected to the other four BBs. This option reduces the C-plane signalling latency but, on other hand, requires additional synchronisation information exchange among BBs.

Second the integration of different ANs via a single WBI requires the definition of a SF in charge of a semantic translation of access specific signalling (e.g. QoS, radio connection quality etc.). This SF might be either located in the AF(s) or in any CN BBs terminating the WBI.

Also related to convergence and access integration is the open issue of defining access independent Connectivity and Mobility state machines.

Another crucial design choice relates to how to enable device slice selection and association. Solving this issue requires a detailed knowledge on how the concept of slicing will be defined at AN. However, assuming AN supporting network slicing, selection and association might be either performed at AN or an CN.

The interface towards other domains (partly indicated as I7 in Figure 5) has to be studied deeper and will not be restricted to the D-plane only. It will have also a C-plane interface and will connect different slices from the same network service provider or different network service provider.

Open and for discussions is the interface towards the verticals, which is not shown in Figure 5. It will be important to express changes in the requirements from the verticals and will offer the possibility to expose a set of information out of the network service infrastructure as defined by the operator.

Finally, in a scenario, where a device may connect simultaneously to multiple slices, some information sharing among slices the device is connected to could lead to significant optimisations or procedure simplifications (e.g. context information sharing, joint inter-slice device location tracking, simplified inter-slice authentication, just to mention a few intuitive ones). Consequently, information or even BBs sharing among different slices becomes a relevant issue to investigate.

# 6 CONCLUSIONS

This paper discussed *Architecture Modularisation* as a key design principle to provide next generation networks with the flexibility required by the diverse set of 5G use cases defined by the vertical business and for a native support of Network Slicing. *Architecture Modularisation* has also been presented as a mean towards heterogeneous access networks integration. Additionally, a modularisation methodology has been presented. Applying such methodology, an overall 5G E2E network architecture has been designed. The architecture includes six fundamental Building Blocks (BB): Access Function (AF), Connectivity Management (CM), Mobility Management (MM), Security and AAA Management (SAM), Flow Management (FM) and Context Generation and Handling Function (CGHF). Each BB groups together a set of mandatory and optional sub-functions, allowing functional customisation on per use case basis. The presented E2E architecture also defines a complete interface model, including inter BB interfaces, C-Plane to D-plane interface, as well as West Bound Interfaces between Core Network and Access Network/Device. Logical architectures of different Network Slices can be defined by a proper interconnection of subsets of (possibly) customised BBs.

The proposed E2E architecture is regarded by the authors as a valid candidate for 5G standards, although some relevant open issues and design choices, highlighted in the discussion, still require some effort and analysis. In the same time, some open issues like the interconnections of the network slices and the interface towards verticals is still under elaborations.

**Acknowledgement**

The authors want to thank the members/partners of the project as well researchers and standardisation delegates in- and outside the project for the very fruitful discussions. The cooperation [3] is not funded by the international and national funding authorities and runs on a self-funding basis.